\documentclass[%
 twocolumn,
 amsmath,amssymb,
 aps, 
 physrev,
floatfix,
]{revtex4-2}

\usepackage{graphicx}
\usepackage{dcolumn}
\usepackage{bm}
\usepackage{tikz}

\begin{document}

\preprint{APS/123-QED}

\title{\textbf{Probe-assisted Depopulation Pumping in Low-pressure Alkali-metal Vapor Cells for Magnetometry } 
}%

\author{M. E. Limes}
 \email{Contact author: limes.mark@gmail.com}
 \altaffiliation[Also at ]{Bradley Department of Electrical and Computer Engineering}
\author{J. Smoot}%
\author{J. Perez}
\author{J. Freeman}
 \author{C. Amano-Dolan}%
\author{D. Peters}
\affiliation{%
 National Security Institute, Virginia Tech, Blacksburg, VA, 24061, USA
}%
\author{W. Lee}
\affiliation{
 Department of Physics, Harvard University, Cambridge, MA, 02138, USA
}
\date{\today}
\begin{abstract}
For precision atomic magnetometry, inert buffer gas is included in alkali-metal vapor cells to significantly broaden hyperfine transitions, which facilitates optical pumping and reduces diffusive relaxation, while also providing non-radiative excited state quenching. 
We show low-buffer gas pressure (below 50 Torr) alkali vapor cells with resolved hyperfine manifolds can also yield high-performance magnetometers. 
For high polarization in $^{87}$Rb, we optically pump $F=2$ states with narrow linewidth $\sigma_+$ light, while tuning a probe beam to depopulate $F=1$ states ($\Delta\nu = 6.8$ GHz from $F=2$). 
The probe tuning then also provides $F=2$ detection with high optical rotation and low probe broadening; we demonstrate top-bottom gradiometry, within a single 25 Torr, 0.5 cc cell, that yields an Earth's field free-precession magnetometer sensitivity of 18 fT/$\sqrt{\text{Hz}}$ with a 1 kHz bandwidth, as well as RF magnetometer sensitivity of 12 fT/$\sqrt{\text{Hz}}$ in a small band about 110 kHz. 
\end{abstract}
\maketitle
Warm alkali-metal vapor systems rival SQUIDs for the world's most sensitive magnetic field measurements \cite{Dang_2010}. 
Warm-atom systems hold several advantages for portable sensing, such as heated operation, rather than the cryogenic cooling and dewars required for SQUIDs. 
Miniature, high-performance atomic sensors are quickly becoming mature, aided by key innovations such as single-mode VCSELs and anodically bonded cells \cite{Kitching_2018}. 
Commercial near-zero field atomic magnetometers show great promise for magnetically shielded magnetoencephalography (MEG) studies involving minor motion \cite{Johnson_2013, Boto_2020, Alem_2023}, using the popular spin-exchange relaxation free (SERF) atomic magnetometer \cite{Allred_2002}. 
However, SERF, and other styles of alkali magnetometers such as free-precession scalar, lose significant sensitivity at Earth’s field and above due to effects that result from hyperfine coupling within the alkali atoms, such as spin-exchange relaxation \cite{Happer_1977} and heading errors \cite{Lee_2021, Hewatt_2025}.

Portable free-precession atomic magnetometers have sufficient performance unshielded in Earth's field to detect MEG signals \cite{Limes_2020, Clancy_2021}, in addition to other various applications, due to their high dynamic range and linearity, resulting from frequency measurements $\omega$ of spins precessing due to a total magnetic field $B$.
Here, calibration is provided by the gyromagnetic ratio $\gamma=\omega/B$ \cite{Sheng_2017, Gerginov_2017, Perry_2020, Gerginov_2020}, rather than the voltage/field measurements of SERF and RF \cite{Savukov_2005} sensors that require frequent calibration. 
Scalar magnetometers also demonstrate high performance within magnetic shielding, comparable to other precision modalities \cite{Li_2011, Sheng_2013, Lucivero_2022}. 
As such, there is high interest in developing methods to improve scalar magnetometer functionality \cite{Hunter_2023}, including introducing field modulations for vector sensing \cite{Alexandrov_2004, Vershovskii_2006, Wang_2025}. 

 For precision magnetometry, it is common that sufficient buffer gas (above $50$ torr N$_2$) is added to vapor cells for optical pumping efficiency, and slowing diffusive wall relaxation \cite{Ottinger_1975,Rotondaro_1997, Romalis_1997, Erhard_2000,Happer_2010, Oelsner_2022}. 
 Here, absorption-line broadening is done for efficiency of both broad and narrow linewidth semiconductor lasers; for the former, broadening causes more absorption line overlap with a wide laser spectrum, and the latter so that all ground hyperfine states can be pumped. 
 Trade-offs of broadening include lowered in-peak absorption efficiency, and lower maximum optical rotation by an incident linearly polarized probe beam \cite{Happer_1967}. 
 We show careful probe tuning can assist optical pumping in low buffer gas pressure vapor cells, leading to high polarizations, while retaining large optical rotation for detection. 
 We give a basic theory, and experimentally demonstrate high sensitivity and high bandwidth for scalar and RF magnetometers.
  Probe-assisted depopulation pumping also suppresses hyperfine coupling effects, such as heading error, and improves tolerance to large fields and inhomogeneous dephasing from magnetic gradients, important for a variety of applications of high-performance atomic magnetometry, such as MEG, magnetic navigation \cite{Bennett_2021}, and nEDM searches \cite{Abel_2020}.

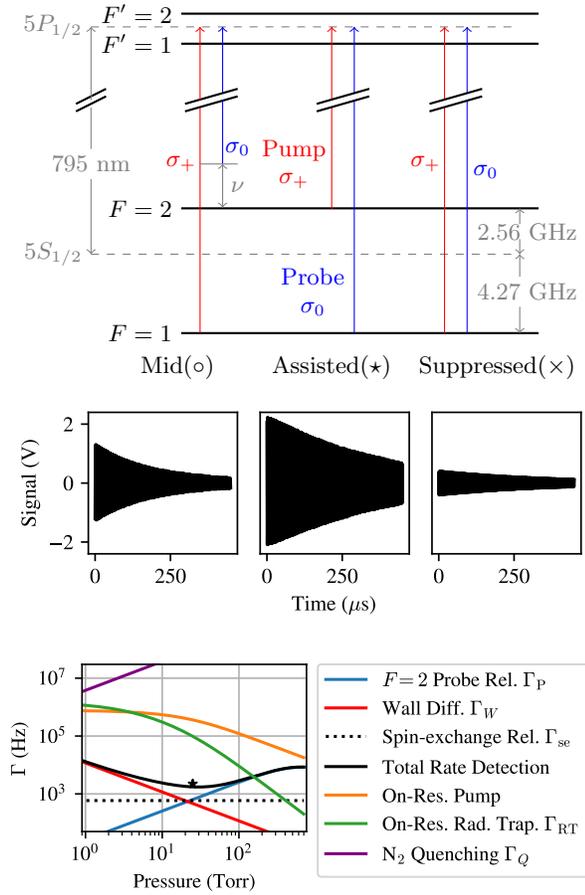
\begin{figure}[tbp]
    \centering
\begin{tikzpicture}[xscale=1, yscale=1]
\def\edde{-1.2}
\def\endde{4.4}
\draw[dashed, gray] (\edde,0) node[left] {5$S_{1/2}$ } -- (\endde,0) ;
\draw[dashed, gray] (\edde,3.024) node[left] {5$P_{1/2}$ } -- (\endde,3.024) ;
\draw[gray, <-] (\edde,0) -- (\edde,1.0);
\draw[gray] (\edde,1.4) -- (\edde,1.95);
\draw[thick] (\edde-0.2,2) -- (\edde+0.2,2.2);
\draw[thick] (\edde-0.2,1.9) -- (\edde+0.2,2.1);
\draw[gray, ->] (\edde,2.05) -- (\edde,3.024);
\node[gray] at (\edde,1.2) {$795$ nm};

\draw[thick] (0,-1.05) node[left] {$F=1$} -- (4.75,-1.05) ;
\draw[<-,gray] (4.5,-1.05) -- (4.5,-0.7);
\draw[->,gray] (4.5,-0.3) -- (4.5,0);
\node[gray] at (4.6,-0.5) {$4.27$ GHz};

\draw[thick] (0, 0.61) node[left] {$F=2$} -- (4.75, 0.61) ;
\draw[<-,gray] (4.5,0) -- (4.5,0.2);
\draw[->,gray] (4.5,0.4) -- (4.5,0.61);
\node[gray] at (4.6,0.3) {$2.56$ GHz};

\draw[thick] (0,2.80) node[left] {$F'=1$} -- (4.75,2.80) ;
\draw[thick] (0,3.20) node[left] {$F'=2$} -- (4.75,3.20);

\def\enPline{2.0}
\def\widToP{0.3}
\draw[red] (\enPline,0.6) -- (\enPline,1.95);
\draw[blue] (\enPline+\widToP,-1.05) -- (\enPline+\widToP,2.05);
\draw[thick] (0.3-0.25,2) -- (1-0.25,2.2);
\draw[thick] (0.3-0.25,1.9) -- (1-0.25,2.1);
\draw[->, red] (\enPline,2.05) -- (\enPline,3.024);
\draw[->, blue] (\enPline+\widToP,2.15) -- (\enPline+\widToP,3.024);
\node[red] at (0,1.2) {$\sigma_+$ };
\node[blue] at (0.75,1.4) {$\sigma_0$};
\node at (\enPline,-1.5) {Assisted($\star$)};

\def\suPline{3.5}
\draw[ red] (\suPline,-1.05) -- (\suPline,1.95);
\draw[ blue] (\suPline+\widToP,-1.05) -- (\suPline+\widToP,2.05);
\draw[thick] (1.8,2) -- (2.5,2.2);
\draw[thick] (1.8,1.9) -- (2.5,2.1);
\draw[->, red] (\suPline,2.05) -- (\suPline,3.024);
\draw[->, blue] (\suPline+\widToP,2.15) -- (\suPline+\widToP,3.024);
\node[red, align=center] at (1.5,1.2) {Pump \\$\sigma_+$ };
\node[blue, align = center] at (1.75,-0.5) {Probe \\$\sigma_0$};
\node at (\suPline+2*\widToP+0.1,-1.5) {Suppressed($\times$)};

\def\noPline{0.25}
\draw[ red] (\noPline,-1.05) -- (\noPline,1.95);
\draw[blue] (\noPline+\widToP,1.2) -- (\noPline+\widToP,2.05);
\draw[thick] (3.3,2) -- (4,2.2);
\draw[thick] (3.3,1.9) -- (4,2.1);
\draw[->, red] (\noPline,2.05) -- (\noPline,3.024);
\draw[->, blue] (\noPline+\widToP,2.15) -- (\noPline+\widToP,3.024);
\node[red] at (3.25,1.2) {$\sigma_+$ };
\node[blue] at (4.05,1.1) {$\sigma_0$};
\node at (\noPline-\widToP,-1.5) {Mid($\circ$)};
\draw[gray] (\noPline,1.2) -- (\noPline+0.5,1.2);
\draw[<->,gray] (\noPline+\widToP,0.62) -- (\noPline+\widToP,1.2);
\node[gray] at (\noPline+0.5,0.9) {$\nu$};

\end{tikzpicture}
 \includegraphics[width=0.45\textwidth]{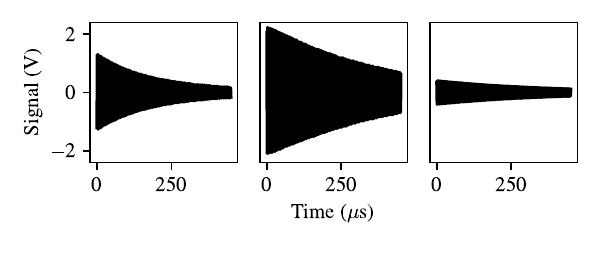}
  \includegraphics[width=0.45\textwidth]{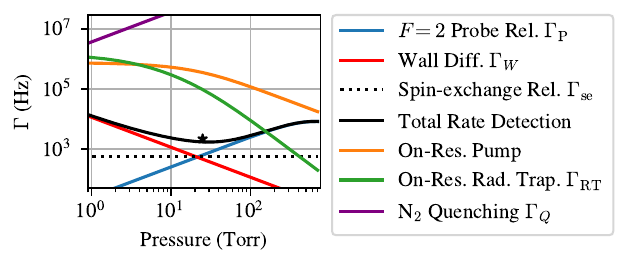}
    \caption{(Top) Grotrian diagram for several regimes, along with representative data. (Bottom) Dominant rates for probe-assisted depopulation pumping; $N_2$ quenching overcomes radiative trapping, with $(\star)$ denoting experimental conditions being wall-relaxation and probe-broadening limited. }
    \label{fig:detuning}
\end{figure}

Scalar free-precession magnetometers are best operated by polarizing alkali atoms to unity into an edge state, maximizing SNR and providing $T_2$ extension by suppressing spin-exchange relaxation \cite{Sheng_2013,Lee_2021}. 
Whether optical pumping occurs pulsed in a plane transverse to the measured scalar magnetic field $B_0$, or is done longitudinally followed with a $\pi/2$ tipping pulse, a resulting transverse spin polarization will freely precesses about a total field $B_0$, where light shifts are avoided with no pump laser during detection periods. 
The probe is tuned to maximize sensitivity, which is a trade-off of maximizing initial optical rotation while reducing probe broadening until it competes with the next largest rate (typically spin-exchange or wall relaxation). 
Increasing $T_2$ is beneficial to optimal sensitivity by roughly the measurement time $T$, which can be seen by converting the Cram\'er-Rao Lower Bound (CRLB) estimate of frequency noise \cite{Hoare_1993, Lucivero_2021} to a field sensitivity
\begin{equation}
\delta B=  \frac{\sqrt{12C}}{2\pi \gamma (A/\rho) T^{3/2}\sqrt{\text{BW}}}~ \left[ \frac{\text{T}}{\sqrt{\text{Hz}}}\right],
\label{eq:crlb}
\end{equation}
with SNR $A/\rho$, integration time $T$, pulsed sensor bandwidth $\text{BW} =1/2T_{\text{rep}}$, and $C$ dependent on $T_2$ and $T$. 
Over a decay period, the average frequency of spin precession is proportional to the magnetic field through the low-field, strong hyperfine coupling gyromagnetic ratio $\gamma$ (e.g.~$^{87}$Rb, $\gamma\!\approx\! 7$ GHz/T). 
To minimize Eq.~\ref{eq:crlb}, a single shot should last $\sim2T_2$ \cite{Limes_2025_2}. 
Measurements are repeated every $T_{\text{rep}}$ for a flat magnetometer bandwidth BW, with caveats---there are amplitude corrections required at higher in-band frequencies, and significant aliasing of out-of-band frequencies \cite{Limes_2025}. 
Artificially higher bandwidth magnetometers may be made by decreasing fitting windows, at a sensitivity loss linearly proportional to increasing bandwidth \cite{Wilson_2020, Jaufenthaler_2021}. 

At low buffer gas pressures for $^{87}$Rb (below $100$ Torr), $F = 2$ and $F = 1$ ground states can be addressed separately by narrow linewidth light \cite{Mcclelland_1986,Dreiling_2012}. 
For this study, we are in a regime where narrow linewidth $\sigma_+$ pump light is efficient, but can only address either $F = 2$ or $F = 1$ at a given time.
As shown by the assisted $(\star)$ regime in Fig.~\ref{fig:detuning}, an additional laser, such as a linearly polarized $\sigma_0$ transverse probe, can be tuned to simultaneously drive transitions from $F=1$.
While the $\sigma_+$ pump drives $F=2$ states to the $m_F = 2$ edge state, an additional depopulation of $F=1$ can achieve a maximum $\times 1.6$ enhancement in edge-stage polarization.  
After a pumping period, the linearly polarized probe tuned to $F=1$ transitions is already conveniently detuned from the highly polarized $F=2$ manifold by the $\Delta\nu = 6.8$ GHz hyperfine splitting, to provide high optical rotation with low magnetic linewidth probe broadening. 
For magnetometry, there are several advantages of probe-assisted depopulation pumping over high buffer gas operation, where the absorption lines of the hyperfine manifolds overlap.
During detection, the probe also continuously depopulates $F=1$, suppressing spin-exchange relaxation at high polarizations even in Earth-scale fields. 
Here, we are wall relaxation and probe broadening limited. 
Also, $F=1$ contributes zero optical rotation, thus no frequency chirp from fast-decaying $F=1$ states (precession $\sim \!1$ kHz difference from $F=2$ at 44 $\mu$T), leaving only non-linear Zeeman heading error effects \cite{Lee_2021, Zhang_2023}. 

A model for probe-assisted depopulation pumping for free-precession magnetometry is made using absorption cross-section $\sigma= \pi r_e c f~\text{Re}[V(\nu)] $ and optical rotation $\phi= l r_e c f n P_x ~\text{Im}[V(\nu)]/2$, with classical electron radius $r_e$, 
speed of light $c$,
$D_1$ oscillator strength $f = 0.34$, path length $l$, alkali number density $n$, and $P_x$ is polarization along the probe axis. 
We use Voigt profiles, as Doppler broadening $\Gamma_G=0.57$ GHz is comparable to buffer-gas broadening $\Gamma_L \approx 18 p/p_0 $ GHz, with buffer gas pressure $p$, and $p_0 = 760$ Torr. 
For narrow laser linewidths, pumping rates are $\Gamma_{\text{P}}= \Phi\sigma$, with photon flux $\Phi = I_0/h\nu$. 
Small cells with very low buffer gas pressures (e.g. $\ll\!1$ cc, 10 Torr) have large diffusive relaxation to walls, depolarizing atoms that can then contribute to radiation trapping \cite{Rosenberry_2007}. 
The radiation trapping rate is estimated by $\Gamma_{\text{RT}} = K(M-1)f_{\text{spon}}\Gamma_{\text{P}}/2$ given in Ref.~\cite{Rosenberry_2007}, where at 100$^{\circ}$C, $K=0.12$, $M = 28$ is the average number of times a photon is emitted before escaping the cell.
We find a wide range of N$_2$ pressures are sufficient to provide non-radiative quenching, with greater than 10 MHz quenching rates $\Gamma_{\text{Q}}$, compared to on-resonance $\sim\!0.5$ MHz pumping/radiation trapping rates. 
For diffusive wall relaxation, we use $\Gamma_{W} = [(\pi/L)^2 + (\mu/R)^2)]D_0 p_o/p$, with $L$ and $R$ cell length and radius,  and diffusion coefficient $D_0$ for Rb in buffer gas \cite{Happer_2010}. 
Shown in Fig.~\ref{fig:detuning}, $\sim$20 torr maximizes coherence time for probe-assisted depopulation pumping for our conditions; we also project a optimum sensitivity at this pressure of $8~ \text{fT}/\sqrt{\text{Hz}}$ with 1.2 kHz bandwidth.
Compare to high buffer gas broadening, where Voigt profiles are replaced by complex Lorentzians and maximum optical rotation occurs with tuning $\nu_{\text{opt}}= \Gamma_L/2$; we also find optimal sensitivity $\delta B$ 
increases as $\Gamma_L$ when probe broadening $\Gamma_{\text{P}}$ competes with buffer gas pressure independent rates $\Gamma_{\text{NP}}$, and bandwidth lowers with $1/\Gamma_L$.

\begin{figure}[t!bp]
\includegraphics[width=0.37\textwidth]{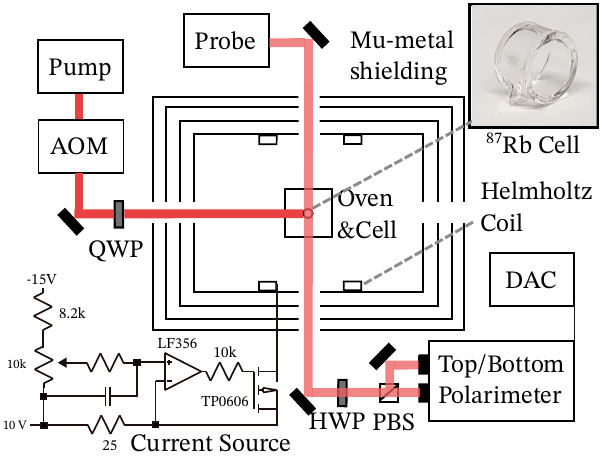}
 \includegraphics[width=0.45\textwidth]{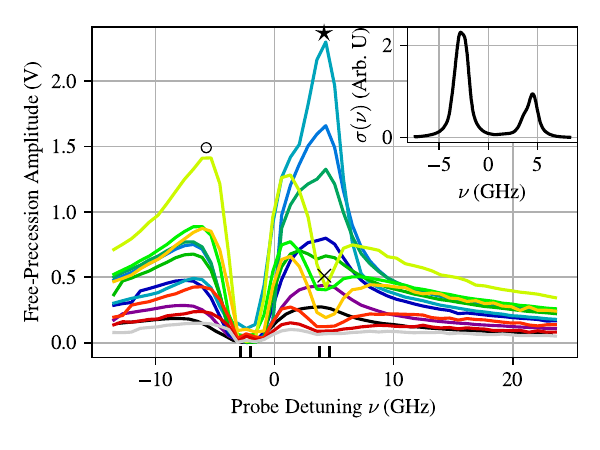}
    \caption{(Top) Schematic of $\sigma_+$ light optically pumping a $^{87}$Rb cell along a 44 {\textmu}T field. After pumping periods, tipping pulses are applied, and spin precession is detected by optical rotation of a linearly polarized probe. 
(Bottom) For a 25 Torr buffer gas $^{87}$Rb cell at 90$^\circ$C, the scalar magnetometer amplitude responds to variation of probe and pump wavelengths. 
The probe detuning is shown along the x-axis, with pump detuning plotted from violet to red (low to high $\nu$) in steps of 1.46 GHz. 
Labeled are mid ($\circ$), assisted ($\star$), and suppressed ($\times$) regimes, and a guide along the x-axis $F\! \rightarrow \!F' : 2\! \rightarrow \! 1, 2 \! \rightarrow \! 2, 1 \! \rightarrow \! 1, 1 \! \rightarrow \! 1$. Inset: Linearly polarized probe absorption cross-section $\sigma(\nu)$. 
}
    \label{fig:detunedData}
\end{figure}

For experiments,  we use a $^{87}$Rb vapor cell with two optical flats separated by 6 mm internal dimension and inner 10 mm diameter (0.47 cc), with 25 Torr of buffer gas, ratio 3:2 Ar:N$_2$. 
The cell is heated in a boron nitride oven with manganin wire driven by 80 kHz CW. 
Both pump and probe are Photodigm 795nm lasers with 0.5 MHz linewidth. 
We use 5-10 mW peak from each laser incident on the vapor cell.
In Fig.~\ref{fig:detunedData}, the pump light is fed into a InterAction Corp.~Acousto-Optic Modulator (AOM), and circularly polarized by a quarter-wave plate (QWP). 
The probe beam passes through the cell's optical flats, and into a half-wave plate (HWP) and balanced polarimeter with polarizing beamsplitter cube (PBS), and a custom top-bottom polarimeter using split photodiodes. 
When balanced, 1 mW of light hits each of four photodiode regions. 
The photon-shot-noise limited signal is digitized with a National Instruments PXI-5922 card. 
To simulate Earth’s field within magnetic shielding, a 40-turn Helmholtz coil is driven by a Libbrecht-Hall circuit \cite{Libbrecht_1993}. 
The bias field and pump are applied along the cylinder axis, with probe transverse. 
After pumping, a resonant $\pi/2$ pulse is applied by a 4 cm square Helmholtz coil, placing spins on the transverse plane.

For a demonstration of probe-assisted depopulation pumping,
 wavelengths of both pump and probe are varied, and a single decay is recorded at 90$^\circ$C. 
Fits to $A\exp(-t/T_2)\sin(\gamma B_0 t+\phi)$ extract amplitudes, which are plotted against probe tuning frequencies $\nu$ in steps of 0.73 GHz, and pump tuning plots vary by color from purple to red, in steps of 1.46 GHz (Fig.~\ref{fig:detunedData}). 
When tuned directly on $F=2$ ground state transitions, the linearly polarized probe is heavily absorbed and undergoes no optical rotation, leaving near-zero signal amplitude. 
For the mid $(\circ)$ regime, we have probe tuning $\nu = -6$ GHz providing a moderate optical rotation signal, with pumping tuned to the $F=1$ ground state transitions. 
When the probe is tuned to 4.27 GHz, there is an effective mechanism to depopulate the $F=1$ manifold to the $F=2$ ground states, such that a pump tuned to -2.56 GHz transitions from the $F=2$ states results in near unity polarization in the $F=2$ manifold, as denoted by $(\star)$. 
Because the probe is already detuned by 6.8 GHz from $F=2$, we also achieve high optical rotation and low probe broadening for detection. 
When the pump tuned to $F=1$ at 4.27 GHz, a probe also tuned for $F=1$ states counteracts any optical pumping to the $F=2$ manifold, resulting in suppression $(\times)$ of signal. 
The response to laser tuning is robust with flipping pump light helicity and field, and remained similar across a range of Earth-scale fields (5 to 150 {\textmu}T).
To confirm cell pressures, the incident linearly polarized probe power is decreased to 0.5 mW, and transmission converted to absorption cross-section $\sigma(\nu)$ by
    $I = I_0 \exp(-n\sigma(\nu)l)$,
with $I$/$I_0$ the output/input light intensity, number density $n$, and $l = 6$ mm path length through cell (Fig.~\ref{fig:detunedData} Inset). 
The absorption cross-section $\sigma(\nu)$ is fit to four Voigt profiles, for each hyperfine ground to excited transition, giving pressure broadening of $\Gamma_B = 0.59 \pm 0.02$ GHz, for $25\pm 1$ Torr of Ar:N$_2$ \cite{Arditi_1964}.

\begin{figure}[tbp]
    \centering
    \includegraphics[width=0.45\textwidth]{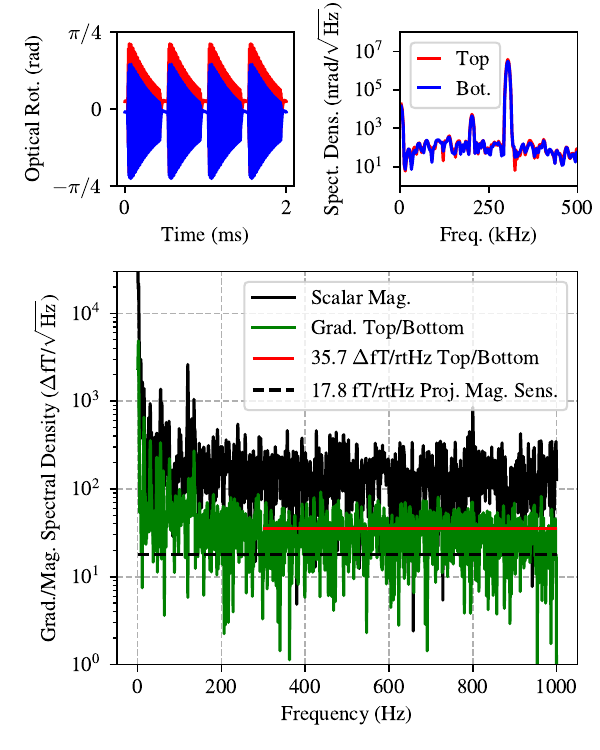}
    \caption{ Top/bottom optical rotation and spectral densities for 0.4 ms shots are shown. Frequencies are extracted each shot for vapor cell top/bottom, and repeated with 2 kHz. A probe-assisted depopulation pumping scalar magnetometer measures 44 {\textmu}T with current supply noise $200$ fT/$\sqrt{\text{Hz}}$.
Top/bottom gradiometry projects magnetic field sensitivity of $17.8\pm 0.3$ fT/$\sqrt{\text{Hz}}$ with 1 kHz bandwidth.}
    \label{fig:scalarmag}
\end{figure}

To demonstrate sensitivity, lasers are tuned for the probe-assisted depopulation pumping regime. 
The probe detection is split to detect vapor cell top and bottom \cite{Kominis_2003} and optical rotation signals are captured, shown in Fig.~\ref{fig:scalarmag}.
The shot-to-shot repetition rate is 2 kHz, with the detection period 0.4 ms each shot and $T_2 = 0.33$ ms. 
Separate shots are fit to decaying sine waves using a non-linear fitting routine, where magnetic fields are extracted by $B = \gamma/\omega$. 
The raw magnetometer performance of roughly $200$ fT/$\sqrt{\text{Hz}}$ is limited by the Libbrecht-Hall supply noise, demonstrated by the roughly $\times5$ lower  gradiometer sensitivity of $35.7\pm0.3$ $\Delta$fT/$\sqrt{\text{Hz}}$ (injecting broadband noise of 1 nT/$\sqrt{\text{Hz}}$ indicates a gradiometer common mode rejection ratio of $\sim\!3000$). 
Gradiometry projects that a magnetometer formed from recombining the cell top and bottom has a sensitivity of 18 fT/$\sqrt{\text{Hz}}$.  
While other methods have sensitivities degrade significantly above Earth's field, see \cite{Limes_2020,Lucivero_2021, Zhang_2023}, we observe no field dependence between 10-100 {\textmu}T. 
Moreover, we retain high sensitivity with large bandwidth using less than $\pi/4$ optical rotation. 
Achieving higher bandwidth by means that do not degrade sensitivity is important for gradient tolerance (and single-shot frequency error requirements), e.g.~from Ref.~\cite{Lucivero_2021}, a similar cell size of $0.5$ cc with high buffer gas, and multi-pass probe, yields 14 fT/$\sqrt{\text{Hz}}$ with 90 Hz bandwidth---our 1 kHz bandwidth requires a factor $\times 10$ less measurement time per shot.
Here gradient relaxation through diffusion is low compared to inhomogeneous $T_2^*$ dephasing, so there is a proportional gain in gradient tolerance with decreasing measurement time $dB/dz = \Delta\phi/(\gamma L T)$, e.g., a signal height loss of $\text{sinc}(\Delta \phi) = 1/e$ results from 160 nT/cm over 0.4 ms, or 16 nT/cm in 4 ms.
The measured gradiometer is higher than the Eq.~\ref{eq:crlb} CRLB estimate of 27.2 fT/$\sqrt{\text{Hz}}$ for observed SNR and $T_2$, with magnetometer of 13.6 fT/$\sqrt{\text{Hz}}$. 
\begin{figure}[ht!p!]
    \centering
    \includegraphics[width=0.45\textwidth]{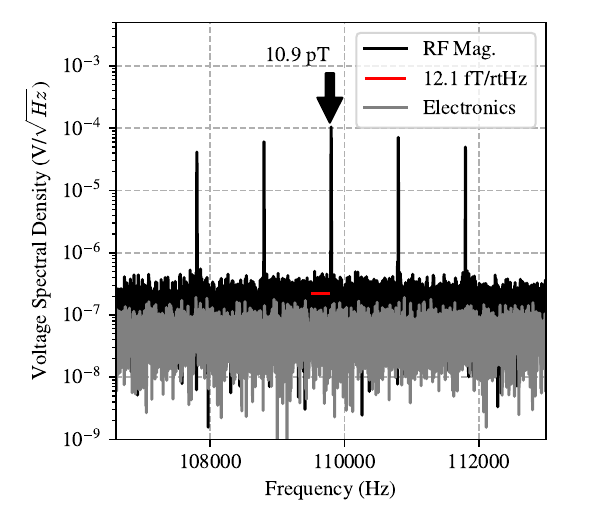}
    \caption{An RF magnetometer is made with the same geometry as the scalar sensor, where the resonant frequency is determined by the 15.7 {\textmu}T scalar field strength. }
    \label{fig:rfMag}
\end{figure}

The same method and geometry can be used for RF magnetometry, which we show experimentally with cell temperature increased to $130^{\circ}$C and continuous probe-assisted depopulation pumping  \cite{Savukov_2005, Savukov_2014, Cooper_2016}. 
To calibrate test fields, the bias field $B_0$ is tuned for resonance near 110 kHz, and Rabi oscillations are driven within the rotating wave approximation where the driving field $B_1$ is much less than $B_0$. 
In Fig.~\ref{fig:rfMag}, we extract a magnetic field sensitivity of $12.1\pm 0.4$ fT/$\sqrt{\text{Hz}}$.
Five separate frequencies detected around 110 kHz are shown with amplitude 10.9 pT, yielding a magnetometer full-width half-max of roughly 3 kHz. 
The spin-projection noise limit is given by 
$\delta B =  (1/\gamma)\sqrt{8/(F_z n V T_2)}$
where $F_z$ is polarization along z, $n$ is number density, and $V$ is volume probed; 
for our parameters we find $\delta B$ is below 10 fT/$\sqrt{\text{Hz}}$.

We demonstrated probe-assisted depopulation pumping for scalar and RF magnetometry in low buffer gas alkali vapor cells.  
The small cell size and low-power, single-pass lasers used give an important proof-of-principle for compact, portable sensor heads using low-pressure vapor cells for improved gradient tolerance and retention of sensitivity in large Earth-scale fields.
The suppression of hyperfine effects opens up a new regime for precision magnetometry, including SERF extensions \cite{Berrebi_2025}; for example, a $^{87}$Rb 50\% polarized $F=2$ manifold can be strongly probed with tuning at $F=1$ ($\Delta\nu = 6.8$ GHz); here the spin evolution due to a magnetic field will only have $F=2$ character. In addition, this method is valid for the other popular alkali metals for precision magnetometry, K and Cs.  
To improve raw sensitivity, a multi-pass configuration may also be used, for spin-noise limited operation \cite{Li_2011, Romalis_2017,Lucivero_2021, Liu_2022, Liu_2023,Yi_2024, Heilman_2024}. 

We thank the Virginia Tech Office of Research and Innovation for support and Brian Saam for discussions.\\
~\\
~\\
~\\
~\\
~\\

\nocite{*}

%

\end{document}